
\newcount\capitolo
\newcount\incremento
\newcount\paragrafo
\font\got=eufm10
 3

\font\bla=msym10
\font\eightbf=cmbx8
\def\gen {\hbox {\got g}}
\def\den{\hbox{\got d}}

\def\ben{\hbox{\got b}}
\def\uen{\hbox{\got u}}
\def\hen{\hbox{\got h}}
\def\nen{\hbox{\got n}}
\def\cen{\hbox{\got c}}

\def\acapo{ {\vskip  0mm plus .2mm } \noindent }
 3
\def\refe#1{\par\bigbreak\centerline{\bf{\bfsmc#1}}
\par\medbreak}
\def\bff#1{{\bf #1}}
\def\bfsmc{\eightbf\bff}
\def\para#1{\global\advance\paragrafo1%
\par\bigbreak\centerline{\bf\number\paragrafo.\ \bfsmc#1}%
\global\incremento=0%
\par\medbreak}
\def\df{\global\advance\incremento by1
{\bf Definition\ \ }}
\def\pf{{\it Proof.\ \ }}
\def\prop{\global\advance\incremento by1 {\bf Proposition
\ \ }}
\def\th{\global\advance\incremento by1 {\bf Theorem
\ \ }}
\def\rem{\global\advance\incremento by1 {\bf Remark
\ \ }}
\def\lem{\global\advance\incremento by1 {\bf Lemma
\ \ }}
\def\cor{\global\advance\incremento by1 {\bf Corollary
\ \ }}

\def\vp{\varphi}

\def\q{\hbox{\bla Q}}

\def\A{{\rm A}}
\def\B{{\rm B}}
\def\C{{\hbox{\bla C}}}
\def\CI{{\rm C}}
\def\D{{\rm D}}
\def\E{{\rm E}}
\def\F{{\rm F}}
\def\L{{\rm L}}
\def\R{{\rm R}}
\def\S{{\rm S}}
\def\I{{\rm I}}
\def\J{{\rm J}}
\def\K{{\rm K}}
\def\H{{\rm H}}
\def\V{{\rm V}}
\def\W{{\rm W}}
\def\U{{\rm U}}
\def\Q{{\rm Q}}
\def\GI{{\rm G}}
\def\M{{\rm M}}

\def\com{{\Delta_{\varphi}}}
\def\pr{^{\prime}}
\def\ppr{^{\prime\prime}}
\def\Y{{\rm Y}}
\def\X{{\rm X}}
\def\T{{\rm T}}
\def\a{{\alpha}}
\def\b{{\beta}}
\def\e{{\varepsilon}}
\def\ga{{\gamma}}
\def\G{{\Gamma}}
\def\l{{\lambda}}
\def\o{{\omega}}

\def\vp{{\varphi}}
\def\d{{\delta}}

\def\vir{,\ldots ,}

\def\lra{{\longrightarrow}}

\def\P {{\rm P}}

\def\cvd{\hfill $\sqcap \hskip-6.5pt \sqcup$}  
\footline={\hss\lower .8cm \hbox{\tenrm\folio}\hss}
\def\sstar{^*}
\hsize=15 true cm
\vsize=20.5 true cm
\baselineskip=.5cm   
\parindent=0pt
\def\interi{\hbox{\bla Z}}

\def\Kl{{\K}_{\l}}
\def\EI{{\E}_{\underline i}}
\def\FI{{\F}_{\underline i}}

\def\g{U_q({\hbox{\got g}})}
\def\gi{U_q^{\varphi}({\hbox{\got g}})}
\def\f{\F_q[{\rm G}]}
\def\fii{\F_q^{\varphi}[{\rm G}]}

\def\gio{\G^{\varphi}({\hbox{\got g}})}

\def\fio{\R_q^{\vp}[\GI]}

\def\gibp{U_q^{\vp}({\hbox{\got b}}_+)}
\def\gibm{U_q^{\vp}({\hbox{\got b}}_-)}

\def\gio{\G^{\vp}({\hbox{\got g}})}
\def\giobp{\G^{\vp}({\hbox{\got b}}_+)}
\def\giobm{\G^{\vp}({\hbox{\got b}}_-)}

\def\fiobp{\R_q^{\vp}[\B_+]}
\def\fiobm{\R_q^{\vp}[\B_-]}
\def\fiobpm{\R_q^{\vp}[\B_{\pm}]}
\def\gaibp{{\overline U}_q^{\vp}({\hbox {\got  b}}_+)}
\def\gaibm{{\overline U}_q^{\vp}({\hbox{\got b}}_-)}
\def\pai{\pi_{\vp}}
\def\pais{{\overline {\pi}}_{\vp}}
\def\ebp{e_{\a}^{\vp}}
\def\ebm{f_{\a}^{\vp}}
\def\eibp{e_i^{\vp}}
\def\eibm{f_i^{\vp}}
\def\tq{\Gamma ({\hbox{\got t}})}
\def\ug{U_{\hbar}({\hbox{\got g}})}
\def\fip{(1+\vp)}
\def\fim{(1-\vp)}
\def\ur{\G_\e^\vp({\hbox{\got g}})}
\def\gr{\F_\e^\vp[\GI]}
\def\rov{{\overline r}}
\def\cov{{\overline c}}
\def\algebra{\A_{\e,\vp}^{(w_1,w_2)}}

 {\bf Multiparameter Quantum Function Algebra}\acapo
 {\bf at Roots of 1}\acapo
\vskip 8mm
{\bf M. Costantini}
\footnote{$^1$}{Dipartimento di Matematica, via Belzoni 7, 35123 Padova,
 Italy \acapo
costantini@pdmat1.unipd.it, fax  39 498758596}
{\bf and M. Varagnolo}
\footnote{$^2$}{Dipartimento di Matematica, via della Ricerca Scientifica,
 00133 Roma,
Italy \acapo varagnolo@vax.mat.utovrm.it, fax  39 672594699}

\vskip 15 mm

\refe{Introduction}
In this paper we consider a multiparameter deformation $\fii$ of the
quantum function algebra associated to a simple algebraic group G.
This deformation has been introduced by Reshetikhin ([R], cf. also [D-K-P1])
and  is constructed from a skew endomorphism  $\vp$
of the  weight lattice of G. When $\vp$ is zero
we get the standard quantum group, that is
the algebra studied by [H-L1-2-3], [Jo] and, in the compact case, by [L-S2].
In the case ($\vp=0$) and when the quantum parameter $q$ is a root of unity,
important results are
contained in [D-L] and [D-P2]. A general $\vp$  has been considered
 in [L-S2] for G compact and $q$ generic.  Here we study the
representation theory at roots of one for a non trivial $\vp$.\acapo
Our arguments are similar to those used by De Concini and Lyubashenko.
Nevertheless there is a substantial difference: when $\vp=0$ the major
 tool is to
understand in detail the $SL(2)$-case which allows to
  construct representations.
Unfortunately there are not multiparameter deformations  for
$SL(2)$.
Moreover the usual right and left actions of the braid group on $\fii$
are not so powerful as in the case $\vp=0$.\acapo
We first (sections {\bf 1.,}{\bf 2.},{\bf 3.})  give
some properties of the multiparameter
quantum function algebra $\gr$ at $\e$, $l$-th root of 1.
 To do this we principally use
 a duality, given in [C-V], between some Borel type subHopfalgebras
of $\fii$. In section {\bf 4.}
we compute  the dimension of the symplectic leaves of G for
the Poisson structure determined by $\vp$.
Our main result  (cor. {\bf 5.7}) is the link between this dimension
 and the dimension of the representations of $\gr$, for "good" $l$.
More precisely, we can see
$\gr$ as a bundle of algebras on G. Its theory of representations
is constant over the T-biinvariant
Poisson submanifold of G (T being the Cartan torus of G) and we have \acapo
\vskip 3mm
{\bf Theorem} {\sl Let $l$ be a "good" integer (see }{\bf 5.5}
{\sl  ) and let $p$
be a point in the
the symplectic leaf $\Theta$ of {\rm G}.
 Then the dimension of any representation of $\gr$
lying over $p$ is divisible by $l^{{1\over 2}dim\Theta}.$}\acapo
\vskip 3mm
Finally, using the results in the first three sections,
we describe esplicitely ({\bf 5.8})
a class of representations of $\gr$.\acapo
\vfill\eject
 \null\vskip-2cm
    Multiparameter Quantum Function Algebra  \hfill
 \vskip11mm
{\bf Notations.}
For the comultiplication in a coalgebra we use the notation
$\Delta x=x_{(1)}\otimes x_{(2)}$. If $H$ is a Hopf algebra, we
denote by $H^{op}$ the same coalgebra with the opposite multiplication
and by $H_{op}$ the same algebra with the opposite comultiplication.\acapo
Let F be a field and let $(\H_i,m_i,\eta_i,\Delta_i,\e_i,S_i),\ i=1,2,$
 be  Hopf algebras. Then an F-linear pairing $\pi  : \H_1\otimes\H_2
 \lra\F$ is called an {\sl Hopf algebra  pairing} [T] if :
 $$\pi (uv\otimes h)=\pi(u\otimes h_{(1)})\pi(v\otimes h_{(2)})
 ,\ \pi(u\otimes hl)=\pi(u_{(1)}\otimes h)\pi(u_{(2)}\otimes l)$$
 $$\pi(\eta_11\otimes h)=\e_2h,\ \pi(u\otimes \eta_21)=\e_1u$$
 $$\pi(S_1u\otimes h)=\pi(u\otimes S_2h),$$
 for $u,v\in\H_1,\ h,l\in\H_2.$
 Moreover $\pi$ is said {\sl perfect}  if it is not degenerate.\acapo
 We denote by $R$ the ring $\q[q,q^{-1}]$ and by $K$ its quotient field
  $\q(q)$. Take a positive integer $l$
  and let  $p_l(q)$ be the $l$th cyclotomic polynomial. We define
 $\q(\e)=\q[q]/(p_l(q))$ ($\e$ being a primitive $l$th root of unity).
 Finally we recall the definition of the $q$-numbers:
 $$(n)_q={{q^n-1}\over{q-1}} ,\ \ (n)_q!=\prod_{m=1}^n{(m)_q} ,\ \
 {\left(\matrix{n\cr m\cr}\right)}_q={{(n)_q!}\over {(m)_q!(n-m)_q!}},$$
 $$[n]_q={{q^n-q^{-n}}\over{q-q^{-1}}} ,\ \ [n]_q!=\prod_{m=1}^n{[m]_q} ,\ \
 {\left[\matrix{n\cr m\cr}\right]}_q={{[n]_q!}\over {[m]_q![n-m]_q!}}.$$

\para{The Multiparameter Quantum Group}
{\bf 1.1.}  Let $\A=(a_{ij})$ be an indecomposable $n\times n$ Cartan matrix;
that is
let $a_{ij}$ be integers with $a_{ii}=2$ and $a_{ij}\leq 0$ for $i\not= j$
and let $(d_1\vir d_n)$ be a fixed $n-$uple of relatively prime positive
integers $d_i$ such that the matrix TA is symmetric and positive definite.
 Here T is the diagonal matrix with entries $d_i$.\acapo
 Consider the free
 abelian group $\P=\sum_{i=1}^{n}{\interi\o_i}$ with basis $\{\o_i|
 i=1\vir n\}$ and define
 $$\a_i=\sum_{j=1}^{n}{a_{ji}\o_j}\ (i=1\vir n),\ \Q=\sum_{i=1}
 ^{n}{\interi\a_i},\ \P_+=\sum_{i=1}^n{\interi_+\o_i};$$
 P and Q are called respectively the weight and the root lattice, the
 elements of $\P_+$ are the dominant weights.\acapo
 Define a bilinear $\interi$-valued pairing on $\P\times\Q$ by the
 rule $(\o_i,\a_j)=d_i\d_{ij}$ ($\d_{ij}$ is the Kronecher symbol);
 it can be extended to symmetric pairings
 $$\P\times\P\lra\interi[{1\over
 {det(\A)}}],\ \q\P\times\q\P\lra\q$$
 where $\q\P=\sum_{i=1}^n{\q\o_i}.$\acapo
 To this setting is associated a complex simple finite dimensional Lie
 algebra ${\hbox{\got g}}$ and a complex connected simply connected simple
 algebraic group G.\acapo
 \vskip 5mm
{\bf 1.2.}  Fix an endomorphism  $\vp$ of the $\q$-vector space $\q\P$ which
 satisfies the following conditions :\acapo
 $$\cases{(\vp x,y)=-(x,\vp y)&$\forall\ x,y\in\q\P$\cr
 \vp \a_i=\d_i=2\tau_i&$\tau_i\in\Q,i=1\vir n$\cr
 {1\over 2}(\vp \l,\mu )\in\interi&$\forall\  \l ,\mu\in\P$\cr}
 \leqno(1.1)$$
We will see later the motivation of the third assumption, now observe
that it implies $\vp\P\subseteq\P.$\acapo
If $\tau_i=\sum_{j=1}^n{x_{ji}\o_j}=\sum_{j=1}{y_{ji}\a_j}$
, let put
 $\X=(x_{ij}),\ \Y=(y_{ij}).$
Then TX is an  antisymmetric matrix and the last two  conditions
in (1.1) are equivalent to the following :
$$\Y\in\M_n(\interi)\cap\T^{-1}\A_n(\interi )\A$$
where $\A_n(\interi )$ denotes the submodule of $\M_n(\interi)$ given
by the antisymmetric matrices.\acapo
The maps
$$1\pm\vp\ :\ \q\P\lra\q\P,\ \a_i\mapsto\a_i\pm\d_i$$
are $\q$-isomorphisms (cf. [C-V]); moreover we have
$$((1+\vp)^{\pm 1}\l,\mu)=(\l,(1-\vp)^{\pm 1}\mu)\ ,\ \forall \l ,\mu\in\P
$$
and so $\fip^\pm\fim^\mp$ are isometries of $\q\P$.
Let us put $r=\fip^{-1},\rov=\fim^{-1}.$\acapo
We like to stress  that if we want to enlarge the results of this paper
to the semisimple case it is enough to ask that $2\A\Y\A^{-1} \in \M_n(
\interi)$, which guarantees $\vp\P\subseteq \P$.\acapo
\vskip 5mm
{\bf 1.3.}
The {\sl multiparameter simply connected quantum group} $\gi$ associated
to $\vp$ ([R],[D-K-P1],cf. [C-V])  is the $K$-algebra on generators
$\E_i,\F_i,\K_{\o_i}^{\pm 1},\
(i=1\vir n)$ with the same relations of the Drinfel'd-Jimbo quantum group
$\g=U_q^0({\hbox{\got g}})$ and with an Hopf algebra structure given by the
following comoltiplication $\com$, counity $\epsilon_{\vp}$ and
antipode $S_{\vp}$ defined on generators $(i=1\vir n; \ \l\in\P)$
$$\cases{\com\E_i=\E_i\otimes\K_{-\tau_i}+\K_{-\a_i+\tau_i}\otimes
\E_i\cr
\com\F_i=\F_i\otimes\K_{\a_i+\tau_i}+\K_{-\tau_i}\otimes\F_i\cr
\com\K_{\l}=\K_{\l}\otimes\K_{\l}\cr},\
\cases{\e_{\vp}\E_i=0\cr
\e_{\vp}\F_i=0\cr
\e_{\vp}\K_{\l}=1\cr},\
\cases{S_{\vp}\E_i=-\K_{\a_i}\E_i\cr
S_{\vp}\F_i=-\F_i\K_{-\a_i}\cr
S_{\vp}\K_{\l}=\K_{-\l}\cr},$$
where for $\l=\sum_{i=1}^n{m_i\o_i}\in\P$ we   use the notation
$\K_{\l}=\prod_{i=1}^n{\K_{\o_i}^{m_i}}.$\acapo
Put $\K_i=\K_{\a_i},\ q_i=q^{d_i}$; we recall the relations in the algebra
$\g$ ([D1],[J]) :
$$\K_{\o_i}\K_{\o_i}^{-1}=1=\K_{\o_i}^{-1}\K_{\o_i},\ \K_{\o_i}\K_{\o_j}=
\K_{\o_j}\K_{\o_i},\leqno(1.2)$$
$$\K_{\o_i}\E_j\K_{\o_i}^{-1}=q_i^{\d_{ij}}\E_j,\ \K_{\o_i}\F_j\K_{\o_i}^{-1}
=q_i^{-\d_{ij}}\F_{j},\leqno(1.3)
$$
$$\E_i\F_j-\F_j\E_i=\d_{ij}{{\K_i-\K_i^{-1}}\over {q_i-q_i^{-1}}}
,\leqno(1.4)$$
$$\sum_{m=0}^{1-a_{ij}}{(-1)^m{\left[\matrix{1-a_{ij}\cr m\cr}\right]}
_{q_i}\GI_i^{1-a_{ij}-m}\GI_j\GI_i^m}=0\ (i\not= j),\leqno(1.5)$$
in the two cases $\GI_i=\E_i,\F_i.$\acapo
\vskip 5mm
{\bf 1.4.}
Let $\gibp$ and $\gaibp$ be the sub-Hopf algebras of $\gi$ generated
 by the $\E_i's\ (i=1\vir n)$ and respectively by the sets
 $\{\K_{\l}|\l\in\P\},\ \{\K_{\l}|\l\in\Q\}.$
Similarly let $\gibm$ and $\gaibm$ be the sub-Hopf algebras of $\gi$
generated by the $\F_i's\ (i=1\vir n)$ and respectively by P and  Q
in the multiplicatively notation of the $\K_{\l}'s.$\acapo
Take an element $u$ in the algebraic closure of $K$ such that
$u^{det(\A+\D)}=q$. Then we know (cf. [C-V]) that the following
bilinear map is a perfect Hopf algebra pairing :
$$\pai\ :\gibm_{op}\otimes\gibp\lra\q(u),\
\cases{\pai (\K_{\l},\K_{\mu})=q^{(r(\l),\mu)}\cr
\pai(\K_{\l},\E_i)=\pai(\F_i,\K_{\l})=0\cr
\pai (\E_i,\F_j)={\d_{ij}\over
{q_i-q_i^{-1}}}q^{(r(\tau_i),\tau_i)}\cr},$$
for $\l,\mu\in\P,\ i=1\vir n.$.
Consider now the antisomorphism $\zeta_{\vp}$
 of Hopf algebras relative to a $\q$-algebra
antisomorphism $\zeta \ : K\lra K,\ q\mapsto q^{-1}$ of
the basic ring, namely
$$\zeta_{\vp}\ :\gi\mapsto U_q^{-\vp}({\hbox{\got g}}),
\ \E_i\mapsto\F_i,\ \F_i\mapsto\E_i,\
\Kl\mapsto\K_{-\l},$$
which send $\gibp$ into $ U_q^{-\vp}({\hbox{\got b}}_-)$ and viceversa.
Then $\pais=\zeta\circ\pi_{-\vp}\circ(\zeta_{\vp}\otimes\zeta_{\vp})$ is
a perfect Hopf algebra pairing :
$$\pais\ :\ \gibp_{op}\otimes\gibm\lra\q(u),\
\cases{\pais(\Kl,\K_{\mu})=q^{-(\rov(\l),\mu)}\cr
\pais(\E_i,\Kl)=\pais(\Kl,\F_i)=0\cr
\pais(\E_i,\F_j)={\d_{ij}\over
{q_i^{-1}-q_i}}q^{-(\rov(\tau_i),\tau_i)}\cr}.$$
\vskip 5mm
For a monomial $\EI=\E_{i_1}\cdots\E_{i_r}$ in the $\E_i's$ and
a monomial $\FI=\F_{i_1}\cdots\F_{i_r}$ in the $\F_i's$ we define
the weight $p(\EI)$ and $p(\FI)$ in the following way :
$$p(\E_r)=p(\F_r)=\a_r,\ p(\EI)=p(\FI)=\sum_{j=1}^n{\a_{i_j}}.
\leqno(1.6)$$
If $v$ is a monomial in the $\E_i$'s or in the $\F_i$'s with
$p(v)=\e$ we will write  $s(v),\ r(v),\ \rov(v),$
instead of  ${1\over 2}\vp(\e),\ {1\over 2}r(\vp(\e)),\ {1\over 2}\rov (
\vp(\e)).$\acapo
The next lemma is proved in [C-V].\acapo
\vskip 5mm
{\bf 1.5.} \lem {\sl For $x$ and $y$ homogeneous polynomials in the
$\E_i$'s and the $\F_i$'s
respectively and for $\l,\ \mu\in\P$ it holds :}
$$\pai(y\Kl,x\K_{\mu})=\pai(y,x)q^{(r(\l),\mu-s(x))-(r(y),\mu)}=
\pi_0(y,x)q^{(r(\l)-r(y),\mu-s(x))},\leqno(i)$$
$$\pais (\Kl x,\K_{\mu} y)=\pais (x,y)q^{(\rov(\l),s(x)-\mu)+
({\overline r}(y),\mu)}={\overline {\pi}}_0(x,y)q^{(\rov(\l)-
{\overline r}(y),s(x)-\mu)}.\leqno(ii)$$
\vskip 5mm
{\bf 1.6.}
Consider the following $R$-subalgebra of $\gi$ :
$$\tq=\{f\in K[\Q]\ |\ \pai(f,\K_{(1-\vp)\l})=\pi_0(f,\Kl )\in R\
\forall \l\in\P\}.$$
In [D-L] is given a $K$-basis $\{\xi_t|\ t=(t_1\vir t_n)\in\interi_+^n\}
$ of $K[\Q ]$ which is an $R$-basis of
$\tq$, namely
$$\xi _t=\prod_{i=1}^n{{\left(\matrix{\K_i;0\cr t_i\cr}\right)}\K_i^
{-[{{t_i} \over 2}]}};\
{\left(\matrix{\K_i;0\cr t_i\cr}\right)}=\prod_{s=1}^t{{{\K_iq_i^{
-s+1}-1}\over{q_i^s-1}}}$$
(for a positive integer $s$, [$s$] denote the integer part).
Note that
$$\{f\in K[(1+\vp)\P]\ |\ \pai (f\otimes\tq)\subseteq R\}=R[(1+\vp)\P],
\leqno (1.7)$$
$$\{f\in K[(1-\vp)\P]\ |\ \pai (\tq\otimes f)\subseteq R\}=
R[(1-\vp)\P].\leqno(1.8)$$
\vskip 5mm
{\bf 1.7.}
Let W be the Weyl group associated to the Cartan matrix A, that is let W be
the finite subgroup of GL(P) generated by the automorphisms $s_i$ of P
given by $s_i(\o_j)=\o_j-\d_{ij}\a_i.$ If $\Omega =\{\a_1\vir \a_n\}$, the
root system corresponding to A is $\Phi=\W\Omega$ while the set of
positive roots is $\Phi_+=\Phi\cap\sum_{i=1}^n{\interi_+\a_i}$.
Fix a reduced expression for the longest element $\o_0$ of W, say
$\o_0=s_{i_1}\cdots s_{i_N}$
and consider the usual total ordering on the set $\Phi_+$ induced by
this choice :
$$\b_1=\a_{i_1}, \ \b_2=s_{i_1}\a_{i_2}\vir\ \b_N=s_{i_1}\cdots s_{i_{N-1}}
\a_{i_N}.$$
Introduced, for $k=1\vir N$, the corresponding root vectors :
$$\GI_{\b_k}=\T_{i_1}\T_{i_2}\cdots\T_{i_{k-1}}(\GI_{i_k}),\
\GI_i=\E_i,\F_i,$$
where the $\T_i$'s are the algebra automorphisms of $\g$ (and so of
$\gi$) introduced by Lustzig up to change $q\leftrightarrow q^{-1},
\ \Kl\leftrightarrow\K_{-\l}$ (see[L2]).\acapo
For a positive integer $s$ define
$$\GI_i^{(s)}={{\GI_i^s}\over{[s]_{q_i}!}},\
\GI_{\b_k}^{(s)}=\T_{i_1}\T_{i_2}\cdots\T_{i_{k-1}}(\GI_{i_k}^{(s)}),$$
always in the two cases $\GI_i=\E_i,\F_i.$\acapo
For $\a\in\Phi_+$ let put
$$q_{\a}=q^{{{(\a,\a)}\over 2}},\ \tau_{\a}={1\over 2}\vp\a ;$$
$$\ebp=(q_{\a}^{-1}-q_{\a})\E_{\a}\K_{\tau_{\a}},\
\ebm=(q_{\a}-q_{\a}^{-1})\F_{\a}\K_{\tau_{\a}};$$
$$\eibp=e_{\a_i}^{\vp},\ \eibm=f_{\a_i}^{\vp}.$$
Note that $\K_{\tau_{\a}}$ commutes with every monomial of weight $\a.$
\acapo
\vskip 5mm
{\bf 1.8.}
Define $\fiobm\pr$ and $\fiobm\ppr$ as the $R$-subalgebras of $\gibp^{op}$
and $\gibp_{op}$ respectively generated by the elements $\ebp,\ \K_{(1-\vp )
\o_i}$ $(i=1\vir n,\ \a\in\Phi_+).$\acapo
Similarly denote by $\fiobp\pr$ and $\fiobp\ppr$ the $R$-subalgebras of $\gibm
^{op}$ and $\gibm_{op}$ respectively generated by the elements $\ebm,\
\K_{(1+\vp)\o_i}$ $(i=1\vir n,\ \a\in\Phi_+).$\acapo
Then, by restriction from $\pai$, we obtain the following two pairings :
$$\pai \pr : \gaibm\otimes_R\fiobm\pr\lra K\ \
\pai\ppr : \fiobp\ppr\otimes_R\gaibp\lra K$$
while by restriction from $\pais$ we get the other two :
$$\pais\pr : \gaibp\otimes_R\fiobp\pr\lra K\ \
\pais\ppr : \fiobm\ppr\otimes_R\gaibm\lra K.$$
We get :
$$\cases{\pai\pr(\F_j,\eibp)=-\d_{ij}\cr\pai\pr(\Kl,\K_{(1-\vp)\mu})=q^{(
\l,\mu)}\cr}\ \ \cases{\pai\ppr(\eibm,\E_j)=\d_{ij}\cr
\pai\ppr(\K_{(1+\vp)\mu},\Kl)=q^{(\mu,\l)}\cr}$$
$$\cases{\pais\pr(\E_j,\eibm)=-\d_{ij}\cr\pais\pr(\Kl,\K_{(1+\vp)\mu})=
q^{-(\l,\mu)}\cr}\ \ \cases{\pais\ppr(\eibp,\F_j)=\d_{ij}\cr
\pais\ppr(\K_{(1-\vp)\mu},\Kl)=q^{-(\mu,\l)}\cr}.$$
\vskip 5mm
We can choose as bases of $\gibp$ and $\gibm$ the elements (see [L2],[D-L]):
$$\xi_{m,t}=\prod_{j=N}^1\E_{\b_j}^{(m_j)}\prod_{i=1}^n
{\left(\matrix{\K_i;0\cr t_i\cr}\right)}\K_i^{-[{{t_i}\over 2}]},\
\eta_{m,t}=\prod_{j=N}^1\F_{\b_j}^{(m_j)}\prod_{i=1}^n
{\left(\matrix{\K_i;0\cr t_i\cr}\right)}\K_i^{-[{{t_i}\over 2}]}$$
\vskip 5mm
{\bf 1.9.} \prop \acapo
$$q^{-\sum_{i<j}(n_i\tau_i,n_j\b_j)}\pai\pr(\eta_{m,t},\prod_{j=N}^1
{(e_{\b_j}^{\vp})^{m_j}\K_{(1-\vp)\l})}=
q^{\sum_{i<j}(n_i\tau_i,n_j\tau_j)}\pai\ppr(\prod_{j=N}^1
{(f_{\b_j}^{\vp})^{n_j}}\K_{(1+\vp)\l},\xi_{m,t})=$$
$${\left (\prod_{i=1}^N\d_{n_i,m_i}q_{\b_i}^{-{{n_i(n_i-1)}\over2}}
\prod_{i=1}^N{\left(\matrix{(\a_i,\l)\cr t_i\cr}\right)}_q q^{-(\a_i,\l)
[{{t_i}\over 2}]}\right)}q^{-\sum_{i=1}^N(n_i\tau_i,\l)}.$$
{\sl Similar formulas hold for $\pais\pr$ and $\pais\ppr$.}\acapo
\vskip 5mm
\pf First of all observe that
$$\com \eibp=\eibp\otimes 1+\K_{-(1-\vp)\a_i}\otimes\eibp,\ \
\com\eibm=\eibm\otimes\K_{(1+\vp)\a_i}+1\otimes\eibm,$$
and, for $\a\in\Phi_+$,
$$\com\ebp=\ebp\otimes 1+\K_{-(1-\vp)\a}\otimes\ebp+e,\ \
\com\ebm=\ebm\otimes\K{(1+\vp)\a}+1\otimes\ebm+f,$$
where $e$ ($f$) is a sum of terms $u_i\otimes v_i$,  $u_i$ and $v_i$ being
linear combination of monomials in the $e_{\b}^{\vp}$ ($f_{\b}^{\vp}$)
and $\Kl$ and $ht(\b)<ht(\a).$ Moreover
$$\pai\pr(\F_{\a},\ebp)=\pai\pr(\F_{\a},e_{\a}\K_{\tau_{\a}})=\pi_0
(\F_{\a},e_{\a})\ \ \forall \ \a\in\Phi_+.$$
Put now $\F=\eta_{n,0},\ \M=\eta_{0,t},\ e^{\vp}=\prod_{j=1}^N{
(e_{\b_j}^{\vp})^{m_j}}$, then
$$\pai\pr(\F\M,e^{\vp}\K_{(1-\vp)\l})=\pai\pr(\F\otimes\M,\com(e^{\vp}
\K_{(1-\vp)\l})=\pai\pr(\F,e^{\vp}\K_{(1-\vp)\l})\pai\pr(\M,\K_{(1-\vp)\l})=$$
$$=\pai\pr(\com\F,\K_{(1-\vp)\l}\otimes e^{\vp})\pi_0\pr(\M,\Kl)=
q^{-(r(\F),(1-\vp)\l)}\pai\pr(\F,e^{\vp})\pi_0\pr(\M,\Kl)=$$
$$=q^{-(s(\F),\l)}\pai\pr(\F,e^{\vp})\pi_0\pr(\M,\Kl).$$
Now, if $e^0=\prod_{j=1}^N(e_{\b_j}^0)^{m_j}$ we have
$$\pai\pr(\F,e^{\vp})=q^{\sum_{i<j}(m_i\tau_i,m_j\tau_j)}
\pai\pr(\F,e^0\K_{\sum_i{m_i\tau_{\b_i}}})=q^{\sum_{i<j}(m_i\tau_i,m_j\tau_j)}
\pi_0\pr(\F,e^0),$$
where the powers of $q$ arises from the commutation of $\K_{\tau_{\a}}$
and the last equality from 1.5.
Since the value of $\pi_0\pr(\F,e^0)$ is calculated in [D-L] (formula
(3.2)) we are done.
\acapo
For the other equality as well as for the case of $\pais$ we proceed
in the same way.\cvd\acapo
\vskip 5mm
{\bf 1.10.}
Define the following $R$-submodules of $\gi$ :
$$\giobp=\{x\in\gaibp\ |\ \pai\ppr(\fiobp_{op}\ppr\otimes x)\subset R\}$$
$$\giobm=\{x\in\gaibm\ |\ \pai\pr(x\otimes{\fiobm\pr}^{op})\subset R\}.$$
It is clear from prop.1.2. that the $\xi_{m,t}$'s and the $\eta_{m,t}$'s
are $R$-bases of $\giobp$ and $\giobm$ respectively and so first of all
they are algebras (cf. [L]) and secondly as algebras they are isomorphic
to $\G^0({\hbox{\got b}}_+)$ and $\G^0({\hbox{\got b}}_-)$ respectively.
 They are also
sub-coalgebras of $\gi$, namely
$$\cases{\com\E_i^{(p)}=\sum_{r+s=n}{q_i^{-rs}\E_i^{(r)}\K_{s(\tau_i-\a_i)}
\otimes\E_i^{(s)}\K_{-r\tau_i}}\cr
\com \F_i^{(p)}=\sum_{r+s=n}{q_i^{-rs}\F_i^{(r)}\K_{-s\tau_i}
\otimes\F_i^{(r)}\K_{r(\a_i+\tau_i)}}\cr
\com {\left(\matrix{\K_i;0\cr t\cr}\right)}=
\sum_{r+s=t}{q_i^{-rs}{\left(\matrix{\K_i;0\cr t\cr}\right)}\otimes
{\left(\matrix{\K_i;0\cr t\cr}\right)}}\cr},\leqno(1.9)$$
(the first two equalities are proved in [C-V], the last in [D-L]).\acapo
\vskip 5mm
{\bf 1.11.}
As a consequence of  1.9.  we get, by restriction,
 two  pairings
 $$\pai\pr : \giobm\otimes_R\fiobm\pr\lra R\ ,\ \pai\ppr : \fiobp\ppr\otimes
 _R\giobp\lra R.$$
 Moreover the same formulas in 1.9. and (1.7), (1.8) give
 $$\{f\in\gibp\ |\ \pai\pr(\giobm_{op}\otimes f)\subset R\}=\fiobm\pr,$$
 $$\{f\in\gibm\ |\ \pai\ppr(f\otimes\giobp^{op})\subset R\}=\fiobp\ppr.$$
 Clearly analogous results hold for $\pais\pr,\ \pais\ppr$ and so we have
 the two perfect pairings
 $$\pais\pr : \giobp\otimes_R\fiobp\pr\lra R\ ,\
 \pais\ppr : \fiobm\ppr\otimes_R\giobm\lra R.$$
 \vskip 5mm
 Most of the definitions and notations introduced up  to now are
 generalisations to the multiparameter
 case of the ones given  in [D-L]. In extending   De Concini-Lyubashenko
 results
we shall only write the parts of the proofs which differ from theirs.\acapo
 \vskip 5mm
 {\bf 1.12.} \lem {\sl The algebras
 $\fiobm\pr,\ \fiobp\pr,\ \fiobp\ppr,\ \fiobm\ppr$ have an Hopf-
 algebra structure for which $\pai\pr,\ \pais\pr,\ \pai\ppr,\ \pais\ppr$
 become perfect Hopf algebra pairings. }\acapo
 \vskip 5mm
 \pf Consider for example $\fiobm\pr$ and let $U_+$ be the sub-K-algebra of
 $\gibp^{op}$ generated by $\{\ebp,\K_{(1-\vp)\l}\ |\ \a\in\Phi_+,\l\in\P\}.$
 We know that (see [L]) the set $\{\eibp,\K_{(1-\vp)\l}\ |\ i=1\vir
n,\l\in\P\}$
 is a generating set for
  $U_+.$ Moreover since
  $$\com\eibp=\eibp\otimes 1+
 \K_{-(1-\vp)\a_i}\otimes\eibp,\ S_{\vp}\eibp=-\K_{-(1-\vp)\a_i}\eibp,
 \ \e_{\vp}\eibp=0,$$
 $U_+$ is an Hopf algebra. So $\com  e \in U_+\otimes U_+$
 for every $e\in\fiobm\pr$.
 In order to   see that indeed $\com e\in\fiobm\otimes\fiobm$
 and to conclude the proof we can proceed as in [D-L](Lemma 3.4).\cvd\acapo
 \vskip 5mm
\para{The Multiparameter Quantum Function Algebra}
{\bf 2.1.} Consider the full subcategory  ${\cal C}_{\vp}$ in $\gi-mod$
consisting of all
finite dimensional modules on which the $\K_i$'s act as powers of $q.$ If
V and W are objects of ${\cal C}_{\vp}$ the tensor product $\V\otimes\W$ and
the dual $\V^*$ are still in ${\cal C}_{\vp}$, namely one can define
$$a(v\otimes w)=\com a(v\otimes w),\ (af)v=f((Sa)v),\ a\in\gi,\ v\in\V,\
w\in\W,\ f\in\V^*.$$ Given$\V\in{\cal C}_{\vp}$, for a vector $v\in \V$ and
a linear form $f\in\V^*$ we define the matrix coefficient $c_{f,v}$ as follows
:
$$c_{f,v} : \gi\lra K,\ x\mapsto f(xv).$$
The $K$-module $\fii$  spanned by all the matrix coefficients is  equipped with
the usual structure of dual  Hopf algebra.
The comultiplication $\Delta$ (which doesn't depend on $\vp$ ) is given by :
$$(\Delta c_{f,v})(x\otimes y)=c_{f,v}(xy),$$
while the multiplication $m_{\vp}$ is given by :
$$m_{\vp}(c_{f,v}\otimes c_{g,w})=c_{f\otimes g,v\otimes w},$$
where $\V,\W\in{\cal C}_{\vp},\ v\in\V,w\in\W,\ f\in\V^*,g\in\W^*,\ x,y\in\gi.$
\acapo
Moreover, since the algebras $\gi$ and $\g$ are equal, in order to
obtain $\fii$ (that as coalgebra is equal to $\F_q^0$[G])
it is enough  to consider the subcategory of ${\cal C}_{\vp}$
given by the highest weight simple modules $\L(\Lambda),\
\Lambda\in\P_+.$\acapo
We
recall that for these modules we have :
$$\L(\Lambda)=\bigoplus_{\l\in\Omega(\Lambda)\subseteq\P}{\L(\Lambda)_{\l}},
\ \L(\Lambda)^*\simeq\L(-\o_0\Lambda),\ \L(\Lambda)^*_{-\mu}=(\L(\Lambda)_
{\mu})^*$$
and that
$$\f=\bigoplus_{\Lambda\in\P_+}{\L(\Lambda)\otimes\L(\Lambda)^*}.$$
\vskip 5mm
{\bf 2.2.}
We want now to link the comultiplication $\com$ in $\gi$ and the multiplication
$m_{\vp}$ in $\fii$ with a bivector $u\in\Lambda^2({\hbox{\got h}})$,
${\hbox{\got h}}$ being the Cartan subalgebra of ${\hbox{\got g}}$ and to do
this we firstly give the Drinfel'd definition of quantized universal
enveloping algebra $\ug$.\acapo
\vskip 5mm
Let $\q[[\hbar]]$ be the ring of formal series in $\hbar$, then $\ug$ is the
$\q[[\hbar]]$-algebra generated, as an algebra complete in the $\hbar$-adic
topology, by the elements $\E_i,\F_i,\H_i\ (i=1\vir n)$ and defining relations
:
$$[\H_i,\H_j]=0,\ [\H_i,\E_j]=a_{ij}\E_j,\ [\H_i,\F_j]=-a_{ij}\F_j$$
added to relations that we can deduce from (1.5) by replacing  $q$ with
$exp({{\hbar}\over 2})$ and $\K_i$ with $exp({{\hbar}\over 2}d_i\H_i).$\acapo
\vskip 5mm
Put now
$$u=\sum_{i,j=1}^n{d_j{\hbar\over 2}u_{ji}\H_i\otimes\H_j}\ \in
\Lambda^2({\hbox{\got
 h}}),$$
 where the matrix $\T\U=(d_iu_{ij})$ is antysimmetric.\acapo
 Then for all $x\in\gi$ using the identity  [R]
 $$exp(-u)(\Delta_0x) exp(u)=\com x$$
 we can compute the $\psi_{ij}$'s, namely
 $$\U=\A^{-1}\X\A^{-1}.$$
Moreover we get the following useful equality (see [L-S2]):
$$m_{\vp}(c_{f_1,v_1}\otimes c_{f_2,v_2})=q^{{1\over 2}((\vp\mu_1,\mu_2)-
(\vp\l_1,\l_2))}m_0(c_{f_1,v_1}\otimes c_{f_2,v_2}),\leqno(2.1)$$
for $\Lambda_i\in\P_+,\ v_i\in\L(\Lambda_i)_{\mu_i},\ f_i\in
\L(\Lambda_i)^*_{-\l_i},\ i=1,2.$\acapo
Observe that (2.1) justifies the condition ${1\over 2}(\vp\l,\mu)\in\interi
,\ \forall\l,\mu\in\P$, required for $\vp$ (see (1.1)).\acapo
\vskip 5mm
{\bf 2.3.}
Since we are interested in the study at roots of 1 we need an integer form
$\fio$ of the multiparameter quantum function algebra. For this purpose
define $\gio$ to be the $R$-subHopf algebra of $\gi$ generated by $\giobp$
and $\giobm$ and
consider the subcategory  ${\cal D}_{\vp}$ of $\gio-mod$ given by the
free $R$-modules of
finite rank in which $\K_i, {\left(\matrix{\K_i;0\cr t\cr}\right)}$ act
by diagonal matrices with eingevalues
$q_i^m,\ {\left(\matrix{m\cr t\cr}\right)}_{q_i}.$
Define $\fio$ as the submodule generated by the  matrix coefficients
constructed
with the objects of ${\cal D}_{\vp}.$
Similarly  define $\fiobp$ and $\fiobm$
starting with opportune subcategories of $\giobp-mod$ and $\giobm-mod$
respectively.\acapo
In completely analogy  with the case $\vp=0$ and essentially in the same way
(cf. prop.4.2 in [D-L])
we can prove that the pairings $\pai\pr,\pais\pr,\pai\ppr,\pais\ppr$ induce
the Hopf algebra isomorphisms
$$\fiobpm\pr\simeq\fiobpm\simeq\fiobpm\ppr\leqno(2.2)$$
and in fact these isomorphisms are the motivations for having introduced
the pairings.
\vskip 5mm
{\bf 2.4.}
Consider now the maps
$$\giobm\otimes_R\giobp \buildrel{\iota_-}\over\lra\gio\otimes_R
\gio\buildrel m\over\lra\gio$$
$$\giobp\otimes_R\giobm\buildrel{\iota_+}\over\lra\gio\otimes_R\gio
\buildrel m\over\lra\gio$$
where $\iota_{\pm}$ are the natural embedding and $m$ is the moltiplication
map.
The corresponding  dual maps composed with the isomorphisms (2.2)
give the injections :
$$\mu_{\vp}\pr : \fio\buildrel \com\over\lra\fio\otimes_R\fio\buildrel
{r_-}\over\lra\fiobm\otimes_R\fiobp\simeq\fiobm\pr\otimes_R\fiobp\pr$$
$$\mu_{\vp}\ppr : \fio\buildrel\com\over\lra\fio\otimes_R\fio
\buildrel{r_+}\over\lra\fiobp\otimes_R\fiobm\simeq\fiobp\ppr\otimes_R
\fiobm\ppr.$$
\vskip 5mm
Let put, for $\M$ in $\tq$, $\l$ in P,
$$\M(\l)=\pai\pr(\M,\K_{(1-\vp)\l})=\pais\pr(\M,\K_{-(1+\vp)\l})=
\pai\ppr(\K_{(1+\vp)\l},\M)=\pais\ppr(\K_{-(1-\vp)\l},\M)=\pi_0
(\M,\Kl).$$
It is now  easy  to prove the following
 (see Lemma 4.3.in [D-L])\acapo
\vskip 5mm
{\bf 2.5.} \lem \acapo
 (i) {\sl The image of $\mu_{\vp}\pr$ is contained in the R-subalgebra
 $\A_{\vp}\pr$ generated by the elements
 $$\ebp\otimes 1,\  1\otimes
 \ebm,\ \K_{\fim\l}\otimes\K_{-\fip\l},\ \l\in\P,\ \a\in\Phi_+.$$}
 (ii) {\sl The image of $\mu_{\vp}\ppr$ is contained in the R-subalgebra
 $\A_{\vp}\ppr$ generated by the elements
 $$1\otimes\ebp,\ \ebm\otimes 1, \
 \K_{-\fip\l}\otimes\K_{\fim\l},\ \l\in\P,\ \a\in\Phi_+.$$}
 \vskip 5mm
 {\bf 2.6.}
Define as in [D-L] the matrix coefficients $\psi_{\pm\l}^{\pm\a}$, that is
for each $\l\in\P_+$  call $v_{\l}$ (resp. $v_{-\l}$)
a choosen  highest  (resp. lowest ) weight vector of $\L(\l)$
(resp. of $\L(-\l)$,
 the irreducible module of lowest weight $-\l$).
Let  $\phi_{\pm\l}$  the unique linear form on $\L(\pm\l)$ such that
$\phi_{\pm\l}v_{\pm\l}=1$ and $\phi_{\pm\l}$ vanishes on the unique
$\tq$-invariant complement of $Kv_{\pm\l}\subset\L(\pm\l)$.\acapo
For $\rho=\sum_{i=1}^n\o_i$, put  $\psi_{\pm\rho}=c_{\phi_{\pm\rho},
v_{\pm\rho}},$ and for $\a\in\Phi_+$ define
$$\psi_\l^\a(x)=\phi_\l(\E_\a xv_\l),\ \psi_\l^{-\a}(x)=
\phi_\l(x\F_\a v_\l),$$
$$\psi_{-\l}^\a(x)=\phi_{-\l}(x\E_\a v_{-\l}),\
\psi_{-\l}^{-\a}(x)=\phi_{-\l}(\F_\a xv_{-\l}).$$
\vskip 5mm
{\bf 2.7.} \prop {\sl The maps $\mu_{\vp}\pr,\ \mu_{\vp}\ppr$
induce algebra isomorphisms}
$$\fio[\psi_{\rho}^{-1}]\simeq\A_{\vp}\pr,\ \
\fio[\psi_{-\rho}^{-1}]\simeq\A_{\vp}\ppr.$$
\vskip 5mm
\pf First of all we specify that what we want to prove is that
the subalgebra generated by $Im(\mu_\vp\pr)$ and $\mu\pr_\vp
(\psi_{\rho}^{-1})$ is indeed $\A_{\vp}\pr$ and similarly for
$\A_{\vp}\ppr.$
Consider the case of $\mu_{\vp}\pr$. First of all we have
$$\mu_{\vp}\pr(\psi_{\rho})=\K_{\fim\rho}\otimes\K_{-\fip\rho}.$$
Moreover  an easy calculation gives
$$ \mu_{\vp}\pr(\psi_{\o_i}^{\a_i})=
-q^{-{1\over 2}(\vp\a_i,\o_i)}\eibp\K_{\fim\o_i}\otimes\K_{-\fip
\o_i},$$
from which
we get $\eibp\otimes 1\in <Im(\mu_{\vp}\pr),\mu_\vp\pr(\psi_\rho^{-1})>$.
\acapo
To see that $\ebp\otimes 1\in
<Im(\mu_{\vp}\pr),\mu_\vp\pr(\psi_\rho^{-1})>$  we
procede as in [D-L], by induction on $ht(\a)$, namely
$$\mu_{\vp}\pr(\psi_\l^\a)=
(-q^{-(\tau_\a,\l)}x(\a,\l)\ebp+d)\K_{\fim\l}\otimes\K_{-\fip\l}$$
where $d$ is a $R$-linear combination of monomials of degree $\a$ in
$e_\b^{\vp}$ with $ht(\b)<ht(\a)$ and
$$x(\a,\l)={{q^{(\a,\l)}-q^{-(\a,\l)}}\over {q_\a-q_\a^{-1}}}.$$
Similar arguments hold  for $1\otimes\ebm$.
\cvd\acapo
\vskip 5mm
\para {Roots of one}
{\bf 3.1.} Consider
a primitive $l$-th root of unity $\e$ with $l$ a positive odd integer
prime to 3 if $\gen$ is of type $G_2$ and define
$\ur=\gio\otimes_R\q(\e),\ \gr=\fio\otimes_R\q(\e),\
\psi : \fio\lra\gr,\ \psi(c_{f,v})={\overline c}_{f,v},$
the canonical projection. By abuse of notations, the image in
$\ur$ of an element of $\gio$ will be indicated with the same symbol.
\acapo
Remark that for $q=l=1$ the quotient of $\G_1^\vp({\hbox{\got g}})$ by
the ideal generated by the  $(\K_i-1)$'s is isomorphic,
as Hopf algebra, to the usual enveloping algebra $U({\hbox{\got g}})$ of
${\hbox{\got g}}$ over the field $\q$; while the Hopf algebra $\F_1^\vp[\GI]$
is isomorphic to the coordinate ring $\q[\GI]$ of G.\acapo
\vskip 5mm
{\bf 3.2.}
It is important to stress some results of Lusztig [L1]
and De Concini-Lyubashenko [D-L]
in the case $\vp=0$ which still
hold in our case principally by virtue of formulas (1.9). More precisely
:\acapo
\vskip 5mm
(i) There exists an epimorphism of Hopf algebras (use (1.9))
$\phi : \gio\lra U({\hbox{\got g}})_{\q(\e)}$ relative to
$R\rightarrow\q(\e)$
such that ($i=1\vir n;\ p>0$) :
$$\phi\E_i^{(p)}=e_i^{({p\over l})},\ \phi\F_i^{(p)}=f_i^{({p\over l})},\
\phi{\left(\matrix{\K_i;0\cr p\cr}\right)}={\left(\matrix{h_i\cr {p\over l}
\cr}\right)}\ ({\rm if\ }l|p,\ 0\ {\rm otherwise});\phi\K_i=1,\ \phi  q=\e.$$
Here $e_i,\ f_i,\ h_i$ are
Chevalley generators for ${\hbox{\got g}}$.
Generators for the kernel $\J$ of $\phi$ are the elements :
$$\E_i^{(p)},\ \F_i^{(p)},\ {\left(\matrix{\K_i;0\cr p\cr}\right)},\
\K_i-1,\ p_l(q)\ (i=1\vir n;\ p>0;\ l\not| \  p).$$
Moreover if $\G_l$ is the free $R$-module with basis
$$\prod_\b\F_\b^{(m_\b)}\xi_t\prod_\a\E_\a^{(n_\a)},\ \ m_\b,t_i,n_\a\equiv
0\ ( mod\ l),$$
then $U({\hbox{\got g}})_{\q(\e)}\simeq \G_l/p_l(q)\G_l.$\acapo
\vskip 5mm
(ii) Denote by I the ideal of $\ur$ generated by $\E_i,\ \F_i,\ \K_i-1\
(i=1\vir n)$. The elements $\prod_\b\F_\b^{(n_\b)}\M\prod_\a\E_\a^{(m_\a)}$,
where M is in the ideal $(\K_i-1|i=1\vir n)\subset\G_\e({\hbox{\got t}})$
or one of the exponents $n_\b,\ m_\a$ is not divisible by $l$, constitute
 an $R$-basis of I. The epimorphism $\phi$ induces the Hopf algebras
 isomorphism $U({\hbox{\got g}})_{\q(\e)}
 \simeq\ur/\I$ and an $R$-basis for
 $U({\hbox{\got g}})_{\q(\e)}$ is given by the elements
$$\prod_\b\F_\b^{(n_\b)}\M\prod_\a\E_\a^{(m_\a)},\ n_\b,m_\a\equiv 0\ (mod\ l)
 ,\ \M\  {\rm  polynomial\ in}\ {\left(\matrix{\K_i;0\cr l\cr}\right)}.$$\acapo
 \vskip 5mm
 {\bf 3.3.}
An important consequence of 3.2. is the existence of  a central
Hopf subalgebra $\F_0$
of $\gr$ which is naturally
 isomorphic to $\q(\e)[\GI].$ An element of $\gr$ belongs to $\F_0$ if and
 only if it vanishes on I and we deduce from [L1] that
$$\F_0=<{\overline c}_{f,v}\  |\ f\in\L(l\Lambda)^*_{-l\nu},
\ v\in\L(l\Lambda)_{l\mu};\
\nu,\mu\in\P_+>,\leqno(3.1)$$
where $ <>$  denotes the  $\q(\e)$-span.\acapo
\vskip 5mm
{\bf 3.4.}
\lem {\sl Let ${\overline c}_{f,v}$ be an element of $\F_0$ and
${\overline c}_{g,w}$ an element of
$\gr$. Then }
$$m_{\vp}({\overline c}_{f,v}\otimes{\overline  c}_{g,w})=
m_0({\overline c}_{f,v}\otimes{\overline c}_{g,w}).$$
\vskip 5mm
\pf It is enough  to consider identity (3.1) and to apply formula
 (2.1).\cvd
\acapo\vskip 5mm
{\bf 3.5.}
\prop  {\sl $\gr$ is a projective module over $\F_0$ of rank $l^{dim \GI}$.}
\acapo
\vskip 5mm
\pf By  3.4. $\gr$ and $\F_\e^0[\GI]$ are the same $\F_0$-modules
and so the result follows from [D-L].\cvd\acapo
\vskip 5mm
{\bf 3.6.} Define $\A_\e^\vp=\A_\vp\ppr\otimes_R\q(\e).$ Let
 $\mu_\e^\vp : \gr\lra\A_\e^\vp$ be the injection induced by $\mu_\vp\ppr;$
 we get the isomorphism (see 2.7.) $\gr[\psi^{-1}_{-l\rho}]\simeq
 \A_\e^\vp.$
Denote by $\A_0^\vp$ the subalgebra of $\A_\e^\vp$ generated by
$$1\otimes(\ebp)^l,\ (\ebm)^l\otimes 1,\ \K_{-\fip(l\l)}\otimes
\K_{\fim(l\l)}\ (\a\in\Phi_+,\ \l\in\P),$$
then $\mu_\e^\vp(\F_0)[\psi^{-1}_{-l\rho}]=\A_0^\vp$ (it is a consequence
of 3.2.,3.3.).\acapo
\vskip 5mm
{\bf 3.7.} A basis for $\A_\e^\vp$ is the following
$$(\F_{\b_N}\K_{\tau_{\b_N}})^{n_N}\cdot\cdot (\F_{\b_1}\K_{\tau_{\b_1}})^{n_1}
\K_{-\fip\o_1}^{s_1}\cdot\cdot\K_{-\fip\o_n}^{s_n}\otimes
\K_{\fim\o_1}^{s_1}\cdot\cdot\K_{\fim\o_n}^{s_n}
(\E_{\b_1}\K_{\tau_{\b_1}})^{m_1}\cdot\cdot
(\E_{\b_N}\K_{\tau_{\b_N}})^{m_N}.$$
Moreover $\A_\e^\vp$ is  a maximal order in its quotient division algebra.
 We can prove this
following the ideas in [D-P1], th. 6.5. (cf also [D-K-P1]). \acapo
\vskip 5mm
{\bf 3.8.} \th {\sl $\gr $ is a maximal order in its quotient division
algebra.}\acapo
\vskip 5mm
\pf In order to repeat the reasoning in  th.7.4. of [D-L] we
need elements $x_1\vir x_r$ in $\F_0$ such that $(x_1\vir x_r)=(1)$
and $\gr[x_i^{-1}]$ is finite over $\F_0[x_i^{-1}].$ In fact, when $\vp\not=
0$ we cannot use left translations (by elements of W) of $\psi_{-l\rho}$.
For $g\in\GI$, let ${\cal M}_g$ be the maximal ideal in $\q(\e)[\GI]$
determined
by it. Then $(\gr)_{{\cal M}_g}$ is a free $(\F_0)_{{\cal M}_g}$-module
of finite type (by 3.5.) and  there exists $x_g\in\F_0\setminus{\cal M}
_g$ (that is $x_g(g)\not= 0$) such that $\gr[x_g^{-1}]$ is a free
$\F_0[x_g^{-1}]$-module  of finite type. Now $\GI=\bigcup_g\D(x_g)$, where
$\D(x_g)=\{x\in\GI|\ x_g(x)\not=0\}$, and so there exist $x_1\vir x_r\in\F_0$
 for which $\GI=\bigcup_{i=1}^r\D(x_i)$, that is the assert. \cvd\acapo
 \vskip 5mm

\para{Poisson structure of G}
{\bf 4.1.} To the quantization $\gio$ of $U(\gen)_{\q(\e)}$ is associated,
in the sense of [D2], a
Manin triple $(\den,\gen,\gen_\vp)$ and a Poisson Hopf algebra structure
on $\F_0=\q(\e)[\GI].$\acapo
The Manin triple is composed of $\gen$, identified with the diagonal
subalgebra of $\den=\gen\times\gen$, and of
$\gen_\vp=\cen_\vp\oplus\uen$, where $\cen_\vp=
\{(-x+\vp(x),x+\vp(x))|\ x\in\hen\},\ \uen=
(\nen_+\times\nen_-),$
$\nen_\pm$ is the nilpotent radical of a fixed  Borel subalgebra $\ben_\pm$ of
$\gen$. Here we denote, by abuse of notation,
  again by $\vp$  the endomorphism of $\hen$
obtained by means of the identification
 $\hen\leftrightarrow\hen^*$ with the Killing form.
 The bilinear form
on $\den$, for which  $\gen$ and $\gen_\vp$ become isotropic Lie subalgebras,
is defined by
$$\prec(x,y),(x\pr,y\pr)\succ=<x,x\pr>-<y,y\pr>,$$
where $<,>$ is the  Killing form on $\gen$.
\acapo
 In order to define a  bracket $\{,\}_\vp$ on $\F_0$ we can
 procede as in [D-L], namely lemma 8.1. still hold after substitution
 $\Delta\leftrightarrow\com$. We want here to give also a direct
 construction  starting
 from the bracket $\{,\}_0$ corresponding to $\vp=0$.\acapo
 \vskip 5mm
 {\bf 4.2.} \prop
{\sl  Let $\Lambda_i\in\P_+,\ v_i\in\L(l\Lambda_i)_{l\mu_i},\ f_i\in
 \L(l\Lambda_i)^*_{-l\l_i},\
 c_i=c_{f_i,v_i},\ i=1,2$ and define $\chi(1,2)=
 {1\over 2}((\vp\mu_1,\mu_2)-(\vp\l_1,\l_2))=-\chi(2,1).$ Then :}
 $$\{\cov_1,\cov_2\}_\vp=\{\cov_1,\cov_2\}_0+2\chi(1,2)m_\vp(\cov_1\otimes
 \cov_2).$$
 \vskip 5mm
 \pf Let $[,]_\vp$ be the commutator in the algebra $\fio.$ The using (2.1)
 we obtain :
 $$[c_1,c_2]_\vp-[c_1,c_2]_0=(q^{l^2\chi(1,2)}-1)m_0(c_1\otimes c_2)-
 (q^{l^2\chi(2,1)}-1)m_0(c_2\otimes c_1).$$
 Now we recall that, by  construction (see [D-L]), if $[c_1,c_2 ]_\vp=
 p_l(q)c$, we put
 $$\{\cov_1,\cov_2\}_\vp=({{p_l(q)}\over {l(q^l-1)}})_{|q=\e}\cov,$$
 and that (by 3.4.) in $\F_0$, $m_\vp$ coincides with $m_0.$
 Then, by projecting in $\gr$ and using the commutativity in $\F_0$,
 we get :
 $$\{\cov_1,\cov_2\}_\vp=\{\cov_1,\cov_2\}_0+(h_{12}-h_{21})
 m_\vp(\cov_1\otimes\cov_2),$$
 where
 $$h_{ij}=({{q^{l^2\chi(i,j)}-1}\over{p_l(q)}}
 \cdot{{p_l(q)}\over{l(q^l-1)}})_{|q=\e}=
 ({{q^{l^2\chi(i,j)}-1}\over{l(q^l-1)}})_{|q=\e}.$$
 Define
 $$p(x)={{x^{l\chi(1,2)}-x^{-l\chi(1,2)}}\over{l(x-1)}}=
 {{x^{-l\chi(1,2)}}\over l}(\sum_{k=0}^{2l\chi(1,2)-1}x^k)\in\q[x,x^{-1}],$$
 then $h_{12}-h_{21}=p(1)=2\chi(1,2)$ and we are done.\cvd\acapo
 \vskip 5mm
{\bf 4.3.} \cor {\sl  (i) Any function $\{\cov_1,\cov_2\}_\vp$, $\cov_i\in
\F_0$, vanishes on the torus T=$exp \hen\subseteq\GI.$}\acapo
{\sl  (ii) Right and left shift by an element of the torus are automorphisms
of the Poisson algebra $\q(\e)[\GI].$}\acapo
\vskip 5mm
\pf  (i) In [D-L] the assert is proved for $\{,\}_0$ then, by 4.2., we only
need to prove that $2\chi(1,2)m_\vp(\cov_1\otimes\cov_2)$ vanishes in the
elements of torus. An easy calculation shows (here we use the identification
$h_i\leftrightarrow{\left(\matrix{\K_i;0\cr l\cr}\right)}$ in agreement
with 3.1.(i)) that, for $t\in\T$,
$(\cov_1\otimes\cov_2)(\com t)\not= 0$ if
$\l_i=\mu_i$, that is if  $\chi(1,2)=0$.\acapo
(ii) The right shift by the element $t\in\T$  is defined as the element
${}^t\cov=\cov_{(1)}\cdot\cov_{(2)}(t)$ (similarly for the left shift) and
then the claim  follows from (i) and from formal properties of
the bracket in a
Poisson Hopf algebra.\cvd\acapo
 \vskip 5mm
 {\bf 4.4.} Let T, $\CI_\vp$, $\U_\pm$, $\B_\pm,$ be the closed connected
 subgroups of G associated to $\hen,\ \cen_\vp,\ \nen_\pm,\ \ben_\pm$ and let D
 be $\GI\times\GI$. Put : $ \GI_\vp=\CI_\vp(\U_+\times\U_-)$,
  $\H=\{(x,x)|\ x\in\T\}$, ${\overline
  {\hen}}=\{(x,x)|\ x\in\hen\}.$ We have the  Bruhat decomposition
  $$\D=\bigcup^{\cdot}_{w\in\W\times\W}{\H\GI_\vp w\GI_\vp}.$$
  The symplectic leaves,  that is the maximal connected
  symplectic subvarieties of G, are the connected components, all isomorphic,
 of
 $\X_w^\vp=p^{-1}(\GI_\vp\backslash\GI_\vp w\H\GI_\vp)$
  for $w$ running in $\W\times\W,$
  where $p :\GI\hookrightarrow\D\rightarrow\GI_\vp\backslash\D$
   is the diagonal immersion followed by the canonical projection (see [L-W]).
   Moreover
   $\X_w^\vp$ are the minimal T-biinvariant Poisson submanifolds  of  G.
   Observe that $\cen_\vp +{\overline{\hen}}=\cen_0+{\overline{\hen}}$
   and so $\CI_\vp\H=\CI_0\H$, that is $\X_w^\vp=\X_w^0=\X_w=
   (\B_+w_1\B_+)\bigcap(\B_-w_2\B_-)$ for all
   $w=(w_1,w_2)\in\W\times\W.$\acapo
 \vskip 5mm
 {\bf 4.5.} \prop  {\sl Let $w=(w_1,w_2)\in\W\times\W.$
 The dimension of a symplectic leaf in  $ \X_{w_1,w_2}$
 is equal to}
 $$l(w_1)+l(w_2)+rk(\fip w_1\fim-\fim w_2\fip),$$
 {\sl where $l(\cdot )$ is the lenght function on $\W$.}\acapo

 \vskip 5mm
 \pf
 Since $p$ is an unramified finite covering of its image, it is enough  to
 calculate the dimension of the $\GI_\vp$-orbits in $\GI_\vp\backslash \D$.
 Moreover $\GI_\vp\subseteq\B_+\times\B_-=\B$, then we can consider the map
$\pi :\GI_\vp\backslash \D\lra\B\backslash\D$,  equivariant for the
right action of $\GI_\vp$ and so preserving $\GI_\vp$-orbits. In $\B\backslash
\D$ the $\GI_\vp$-orbits coincide with the B-orbits which are equals to
$\B\backslash(\B_+w_1\B_+\times\B_-w_2\B_-)=\Theta(w_1,w_2).$
Note that  $\pi$ is a principal $\T/\G$-bundle,
where $\G=\{t\in\T|\ t^2=1\}.$ Let $\Theta$ be a $\GI_\vp$-orbit in D
such that $\pi(\Theta)=\Theta(w_1,w_2)$, then
$\pi_{\mid\Theta} : \Theta\lra\Theta(w_1,w_2)$ is a
principal $\T_{w_1,w_2}/\G$-bundle where
$\T_{w_1,w_2}=\{t\in\T|\ t\Theta=\Theta\}.$
 From it follows $dim\Theta=dim(\T_{w_1,w_2}/\G)+dim\Theta(w_1,w_2)$,
that is
$$dim \Theta=dim\T_{w_1,w_2}+l(w_1)+l(w_2).$$
In order to calculate $dim\T_{w_1,w_2}$ take $n_1, n_2$ representatives
of $w_1, w_2$ in the normalizer of T. We get $t\Theta=\Theta$ if and only
if there exist $(t_1,t_2), (s_1,s_2)\in\CI_\vp$ and
$(s_1,s_2)(t,t)=(n_1,n_2)(t_1,t_2)(n_1,n_2)^{-1}.$
Let $u,v$ be elements in $\hen$ such that
$$(s_1,s_2)=(exp(-u+\vp u),exp(u+\vp u)),\ \ (t_1,t_2)=(exp(-v+\vp v),
exp(v+\vp v)).$$
 We are so reduced to find $x\in\hen$ for which
 $$\cases{-u+\vp u+x=w_1(-v+\vp v)\cr
 u+\vp u+x=w_2(v+\vp v)\cr},$$
 that is
 $$\cases{2x+2\vp u=(-w_1\fim +w_2\fip) v\cr
 2u=(w_1\fim+w_2\fip)v\cr}.$$
 We find $2x=(\fip w_1\fim-\fim w_2\fip)v$ and so
 $$dim\T_{w_1,w_2}=rk(\fip w_1\fim -\fim w_2\fip).
 $$\cvd\acapo

\para {Representations}
{\bf 5.1.}  In all this paragraph we shall substitute the basic field $\q(\e)$
with $\C$.
The fact that $\gr$ is a projective module of rank
$l^{dim(\GI)}$ over $\F_0$ allows us to define a bundle of algebras
on G with fibers $\gr(g)=\gr/{\cal M}_g\gr$ (for more details
on this construction confront  section 9 in [D-L]). From the results
of previous chapter also in our case the algebras $\gr(g)$ and
$\gr(h)$ are isomorphic for $g,h$ in the same $\X_{w_1,w_2}$ that is,
 using the central character map $Spec(\gr)\lra Spec(\F_0)=\GI$,
 the representation theory of $\gr$ is constant on the sets $\X_{w_1,w_2}.$
 \acapo\vskip 5mm
{\bf 5.2.} Let $w_1,w_2$  be two elements in W.
Choose reduced expressions for
them, namely $w_1=
s_{i_1}\cdots s_{i_t}$, $w_2=s_{j_1}\cdots s_{j_m}$, and consider
the corresponding  ordered sets of positive roots
$\{\b_1\vir\b_t\}$ and $\{\ga_1\vir\ga_m\}$ with $\b_1=\a_{i_1}$,
 $\b_r=s_{i_1}\cdots s_{i_{r-1}}\a_{i_r}$ for
 $r>1$ and similarly for the $\ga_i$'s.
 Define $\algebra$ as the subalgebra in $\A_\e^\vp$ generated
 by the elements
 $$1\otimes e_{\b_i},\ f_{\ga_j}\otimes 1,\ \K_{-\fip\l}
 \otimes\K_{\fim\l}\ \ (i=1\vir t,\ j=1\vir m,\ \l\in\P),$$
 and  put $\A_{0,\vp}^{(w_1,w_2)}=\algebra\cap\A_0^\vp$.
Note that these definitions do not  depend on the reduced
expressions (see [D-K-P2]). The algebra $\algebra$ is a free module of
rank $l^{l(w_1)+l(w_2)+n}$ over its central subalgebra $\A_{0,\vp}
^{(w_1,w_2)}$ and so it is finite over its centre and has finite
degree. We will call $d_\vp(w_1,w_2)$ the degree of $\algebra$.
\acapo\vskip 5mm
{\bf 5.3.} There is an algebra isomorphism  $\A_{0,0}^{(w_1,w_2)}\simeq
 \A_{0,\vp}^{(w_1,w_2)}$ induced by the  isomorphism between the algebras
 $\A_0^0$ and $\A_0^\vp$ given by
 $$1\otimes e_\a^l\mapsto1\otimes(\ebp)^l,\
 f_\a^l\otimes 1\mapsto (\ebm)^l\otimes 1,\ \K_{-l\l}\otimes
 \K_{l\l}\mapsto \K_{-\fip l\l}\otimes\K_{\fim l\l}.$$
 Therefore $Spec(\A_{0,\vp}^{(w_1,w_2)})$ is birationally isomorphic to
 $ \X_{w_1,w_2}\bigcap Spec (\A_0^\vp)$ (cf prop. 10.4 in [D-L]). From this,
 and
 reasoning as in [D-L], it follows that the dimension of any representation of
 $\gr$ lying over a point in $\X_{w_1,w_2}$ has dimension divisible
  by $d_\vp(w_1,w_2).$\acapo
\vskip 5mm
{\bf 5.4.} In order to calculate the degree $d_\vp(w_1,w_2)$ we introduce
another set of generators
for $\algebra.$
Call $\Xi$ the antisomorphism of algebras $\Xi :\gi\lra\gi$ which
is the identity on $\E_i,\F_i,q$ and send $\K_i$ into $\K_i^{-1}$; we
get  $T_i^{-1}=\Xi\T_i\Xi$ (see [L2]). For $\a\in\{\gamma_1\vir
\gamma_m\}$, let  $\Xi( f_\a^0)\K_{\tau_\a}=
{\ebm}\pr$.
Observe that, for $r=1\vir m$, $\Xi(\F_{\gamma_r}) =\T_{i_1}^{-1}\cdots
\T_{i_{r-1}}^{-1}(\F_{i_r}).$
We want now to show that the sets $\{f_{\b_i}^\vp,
|\ i=1\vir m
\}$ and $\{{f_{\b_i}^\vp}\pr,
|\ i=1\vir m\}$
generate the same subalgebra of $\fiobp\ppr$. Let $\H_1,\H_2$ be the
subalgebras respectively generated by these sets. From
$\Xi(U_q^\vp(\nen_-))=U_q^\vp(\nen_-)$ and $\T_i^{\pm 1}(\fiobp\ppr)
\subseteq \fiobp\ppr$ for every $i$, follow that
${f_\a^0}\pr$ belongs to the algebra generated by $\{f_{\gamma_i}
^0|i=1\vir m\}$ and
$${\ebm}\pr={f_\a^0}\pr\K_{\tau_\a}=
(\sum_s{c_s (f_{\gamma_m}^0)^{s_m}\cdots(f_{\gamma_1}^0)^{s_1}})\K_{\tau_\a}=
\sum_s{c\pr_s(f_{\gamma_m}^\vp)^{s_m}\cdots(f_{\gamma_1}^\vp)^{s_1}}
\in\H_1,$$
where $c_s\pr=q^{r_s}c_s$ for an integer $r_s$. In a similar way we can show
that $\ebm\in\H_2.$
\acapo
Put now, in $\algebra$,
$$x_i\pr=1\otimes e_{\b_i}^\vp\ (i=1\vir t),\
y_j\pr= {f_{\gamma_j}^\vp}\pr\otimes 1\ (j=1\vir m),\
z_r\pr=\K_{-\fip\o_r}\otimes\K_{\fim\o_r}\ (r=1\vir n).$$
As in the case $\vp=0$, $\algebra$ is an iterated twisted polynomial algebra
and the corresponding quasipolynomial algebra is generated by elements
$x_i,\ y_j,\ z_r$ with relations
which are easily found (see [L-S1] and [D-K-P2]). Namely :
$$x_ix_j=\e^{(\b_i,\fip\b_j)}x_jx_i\ (1\leq j<i\leq t),\
y_iy_j=\e^{-(\gamma_i,\fip\gamma_j)}y_jy_i\ (1\leq j<i\leq m),$$
$$z_iz_j=z_jz_i\ (1\leq i,j\leq n),\ x_iy_j=y_jx_i\ (1\leq i\leq t,
i\leq j\leq m),$$
$$z_ix_j=\e^{(\fim\o_i,\b_j)}x_jz_i\ (1\leq i\leq n,i\leq j\leq t),\
z_iy_j=\e^{(\fip\o_i,\gamma_j)}y_jz_i\ (1\leq i\leq n,i\leq j\leq m).$$
\vskip 5mm
{\bf 5.5.} Let $\interi_\vp$ be the
ring $\interi[(2d_1\cdots d_n det\fim)^{-1}]$
and  denote by $\vartheta$ the isometry $\fip\fim^{-1}$.
For each
pair $(w_1,w_2)$ in $\W\times\W$, consider the map $e_\vp(w_1,w_2)=
1-w_1^{-1}\vartheta^{-1}w_2\vartheta :\P\otimes_{\interi}\q\leftrightarrow
\Q\otimes_{\interi}\q.$ Define $l(\vp)$ to be the least positive integer for
which, for every $(w_1,w_2)$, the image of $\P\otimes\interi_\vp[l(\vp)^{-1}]$
is a split summand of
 $\Q\otimes\interi_\vp[l(\vp)^{-1}]$ (in special cases, namely
when $\vartheta$ fix the set of roots, one can esplicitely take $l(\vp)=
a_1\cdots a_n$, where
 $\sum_{i=1}^n{a_i\a_i}$ is the longest root, as in [D-K-P2],
but in general we need a case by case analysis).\acapo
An integer $l$ is said to be
 a $\vp${\sl -good integer} if, besides being prime to the
$2d_i$, it is prime to $det\fim$ and $l(\vp)$.

\vskip 5mm
{\bf 5.6.} \th {\sl Let $l$ be a $\vp$-good integer, $l>1$. Then,}
$$d_\vp(w_1,w_2)=l^{{1\over 2}(l(w_1)+l(w_2)+rk(\fim w_1\fip-\fip w_2\fim
)}.$$
\vskip 5mm
\pf   We work over $\S= \interi_\vp[l(\vp)^{-1}]$. Let  $w_1,w_2$ be in W.
 Consider free $\S$-
modules $\V_{w_1}$,  $\V_{w_2}$ with basis $u_1\vir u_t$ and
$ v_1\vir v_m$ respectively.
Define on $\V_{w_1}$,  $\V_{w_2}$ skew symmetric
bilinear forms by
$$<u_i|u_j>=(\b_i,\fip\b_j)\ (1\leq j<i\leq t),\
<v_i|v_j>=(\gamma_i,\fip\gamma_j)\ (1\leq j<i\leq m),$$
and denote by $C_{w_1}^\vp$,  $C_{w_2}^\vp$ their  matrices in the bases
of the $u_i$'s and  $v_j$'s respectively. Finally let $D_{w_1}^\vp,
$ $D_{w_2}^\vp$ be  the
$t\times n,\ m\times n$ matrices whose entries are $(\b_i,\fim\o_j)$ and
$(\gamma_i,\fip\o_j)$ respectively. Put
$$\Delta_{w_1,w_2}^\vp={\left(\matrix{
C_{w_1}^\vp&0&D_{w_1}^\vp\cr
0&-C_{w_2}^\vp&D_{w_2}^\vp\cr
-{}^tD^\vp_{w_1}&-{}^tD_{w_2}^\vp&0\cr}\right)},\
 \Delta={\left(\matrix{C^0_{w_1}&0&D^\vp_{w_1}\cr
0&-C^0_{w_2}&D^\vp_{w_2}\cr-{}^tD_{w_1}^\vp&-{}^tD_{w_2}^\vp&0\cr}
\right)}.$$
We want first of all prove that
$\Delta^\vp_{w_1,w_2}$ is equivalent to $\Delta,$
 that is we want to exibite  an $n\times t$ matrix
$M_1$ and an $m\times n$ matrix $M_2$ for which  :
$${\left (\matrix{1&0&0\cr0&1& M_2\cr0&0&1\cr}\right)}
\Delta_{w_1,w_2}^\vp
{\left(\matrix{1&0&0\cr0&1&0\cr  M_1&0&1\cr}\right)}
=\Delta,$$
or equivalently for which
$$C_{w_1}^\vp+D_{w_1}^\vp M_1=C^0_{w_1},\
C_{w_2}^\vp+M_2({}^tD_{w_2}^\vp)=C^0_{w_2},\
D_{w_2}^\vp M_1=M_2({}^tD_{w_1}^\vp).$$
First of all we need some notations. If $f:\V_1\rightarrow\V_2$ is a linear
map  and ${\cal B}_1$ (resp. ${\cal B}_2$) is a basis
of $\V_1$ (resp. $\V_2$) we will indicate by $\M(f,{\cal B}_1,{\cal B}_2)$
 the matrix of $ f$ in these given bases. Let now $\check{\a}_i={{\a_i}
 \over {d_i}}$ and denote by  $\nu:\S
 \P\lra(\S\P)\sstar$ the map given by $\nu(
 \check{\a}_i)=\o_i\sstar$ or, equivalently, the map which send $\a_i$ to the
 linear form $(\a_i,\cdot)$. We define the following maps :
 $$c_1^\vp:\V_{w_1}\lra\V_{w_1}\sstar,\ u_j\mapsto \sum_{i>j}{(\b_j,
 \fip\b_i)u_i\sstar}
 -\sum_{i<j}{(\b_i,\fip\b_j)u_i\sstar};$$
 $$c_2^\vp:\V_{w_2}\lra\V_{w_2}\sstar,\ v_j\mapsto \sum_{i>j}{(\ga_j,
 \fip\ga_i)v_i\sstar}
 -\sum_{i<j}{(\ga_i,\fip\ga_j)v_i\sstar};$$
 $$d_1^\vp:\interi\pr\P\lra\V_{w_1}\sstar,\
 \o_i\mapsto\sum_j{(\fip\b_j,\o_i)u_j\sstar};\
 d_2^\vp:\interi\pr
 \P\lra\V_{w_2}\sstar,\ \o_i\mapsto\sum_j{(\fim \ga_j,\o_i)v_j
 \sstar};$$
 $$h_1:\V_{w_1}\lra\S\P,\ u_j\mapsto\b_j;\ h_2:\V_{w_2}\lra
 \S\P
 ,\ v_j\mapsto\ga_j.$$
 Then we have
 $$d_1^0=\nu h_1,\ d_2^0=\nu h_2,\ d_1^\vp=d_1^0\fim,\
 d_2^\vp=d_2^0\fip,$$
 and we can easy verify that
 $$c_1^\vp-c_1^0=-d_1^0\vp h_1,\ c_2^\vp-c_2^0=-d_2^0\vp
 h_2.$$
 Moreover, if $Z=\M(\vp,\{\o_i\} ,\{\o_i\})$, we get
 $$C_{w_1}^\vp=\M(c_1^\vp,\{u_i\},\{u_i\sstar\}),\
 C_{w_2}^\vp=\M(c_2^\vp,\{v_i\},\{v_i\sstar\});$$
 $$D_{w_1}^\vp=\M(x_1^\vp,\{\o_i\},\{u_i\sstar\})=D_{w_1}^0(1-Z),\
 D_{w_2}^\vp=\M(x_2^\vp,\{\o_i\},\{v_i\sstar\})=D_{w_2}^0(1+Z).$$
Let
 $R=\M(id,\{\check{\a}_i\},\{\o_i\})$ and define
 $$M_1=(1-Z)^{-1}ZR({}^tD^\vp_{w_1}),\ M_2=D^\vp
 _{w_2}ZR(1+{}^tZ)^{-1},$$
now it is  a straightforward computation to verify that these two matrices
satisfy the required properties.\acapo
Let now $d$ be the the map corresponding to $\Delta$ with respect
to the bases $\{u_i,v_j,\o_r\}$ and
$\{u_i^*,v_j^*,\check{\a}_r\}.$
We want to show that the image of $d$ is a
split direct summand, and to calculate
the rank of $\Delta.$
 The result will then follow from the proposition on page 34 in
[D-P1], due to restrictions imposed to $l$.
 From the results in [D-K-P2] we now that
the map corresponding to
 $(C_{w_1}^0,D_{w_1}^\vp):\V_{w_1}\oplus\S\P\lra\V_{w_1}$
is surjective with kernel
 $$\{(u_{\fim\l},(1+\fim^{-1}w_1\fim)\l)|\ \l\in\S\P\}.$$
Similarly the map corresponding to
$(-C_{w_2}^0,D_{w_2}^\vp):\V_{w_2}\oplus\S\P\lra\V_{w_2}$ is surjective
with kernel  $$\{(-v_{\fip\l},(1+\fip^{-1}w_2\fip)\l)|\ \l\in\S\P\}.$$
Here, for $\l=\o_r\ (r=1\vir n)$, we have defined
 $$\I^1_r=\{k\in\{1\vir t\}|\ i_k=r\},\ \I_r^2=\{k\in\{1\vir m\}|\
 j_k=r\};\
 u_\l=\sum_{k\in\I_r^1}{u_k},\  v_\l=\sum_{k\in\I^2_r}{v_k}$$
and  the extension of  the definition
of  $u_\l,\ v_\l$ to all $\l\in\P$ is the only compatible with linearity.
We start the study of $d$. We consider the first row
$(C_{w_1}^0,0,D_{w_1}^\vp):\V_{w_1}\oplus\V_{w_2}\S\P\lra\V_{w_1}$;
it is surjective with kernel
 $$\H=\{(u_{\fim\l},v,(1+\fim^{-1}w_1\fim)\l)|\ \l\in\S\P\,\ v\in\V_{w_2}\}.$$
 Our aim is now to study the image of the restriction of $d$ on H. We proceed
 as follows. We define the composite map $f:\V_{w_2}\oplus\S\P\lra\H\lra
 \V_{w_2}^*\oplus\S\Q,$
 $$(v,\l)\mapsto(u_{\fim\l},v,(1+\fim^{-1}w_1\fim)\l)\mapsto
 d(u_{\fim\l},v,(1+\fim^{-1}w_1\fim)\l).$$
 With respect to the bases $\{v_i,\o_j\}$, $\{v_i^*,
 \check{\a}_j\}$, $f$ is represented by the matrix
 $${\left(\matrix{-C_{w_2}^0&D_{w_2}^\vp(1+(1+\Phi)^{-1}W_1(1-\Phi))\cr
 -{}^tD_{w_2}^\vp&-(1+\Phi)(1-W_1)(1-\Phi)\cr}\right)},$$
 where $W_1,\Phi$ are the matrices representing
$w_1,\vp$ respectively with respect
 to the basis $\{\o_j\}.$
To study this matrix is equivalent to study its opposite
 transpose (since we are essentially interested in their elementary divisors).
  Let $M=(1+(1-\Phi)^{-1}W_1(1-\Phi))$,
 $N=(1+\Phi)(1-W_1)(1-\Phi).$ We consider
  therefore the matrix
  $${\left(\matrix{-C_{w_2}^0&D_{w_2}^\vp\cr
  -{}^tM{}^tD_{w_2}^\vp&{}^tN\cr}\right)}.$$
  Let $g:\V_{w_2}\oplus\S\P\lra\V_{w_2}^*\oplus\S\Q$ be the map represented
  by this matrix with respect to the bases
 $\{v_i,\o_j\}$, $\{v_i^*,\check{\a}_j\}.$
  Then $(-C^0_{w_2},D_{w_2}^\vp):\V_{w_2}\oplus\S\P\lra\V_{w_2}^*$ is
  surjective with kernel
  $$\L=\{(-v_{\fip\l},(1+\fip^{-1}w_2\fip)\l)|\ \l\in\S\P\}$$
  and we are left to study the following composite $e: \S\P\lra\L\lra\S\Q$,
  $$\l\mapsto(-v_{\fip\l},(1+\fip^{-1}w_2\fip)\l)\mapsto
  g((-v_{\fip\l},(1+\fip^{-1}w_2\fip)\l).$$
  With respect to the bases $\{\o_i\}$ and $\{\check{\a}_i\}$, $e$
  is represented by the matrix
  $${}^tM(1-\Phi)(1-W_2)(1+\Phi)+{}^tN(1+\Phi)^{-1}
  W_2(1+\Phi)),$$ that is
  $$(1+{}^t(1-\Phi){}^tW_1{}^t(1-\Phi)^{-1})(1-\Phi)(1-W_2)(1+\Phi)+
  {}^t(1-\Phi)(1-{}^tW_1){}^t(1+\Phi)(1+(1+\Phi)^{-1}W_2(1+\Phi)).$$
  Since $w$ is an isometry and $\vp$ is skew (one should use at each
  step appropriate bases), we get that $e(\l)$ is the element
  $$((1+\fip w_1^{-1}\fip^{-1})\fim(1-w_2)\fip+
  \fip(1-w_1^{-1})\fim(1+\fip^{-1}w_2\fip)\l,$$ that is
  $$e(\l)=(1+\fip w_1^{-1}\fip^{-1})\fim(1-w_2)\fip+
  \fip(1-w_2^{-1})\fim(1+\fip^{-1}w_2\fip).$$
  It follows that
  $$e(\l)=2\fip\fim-2\fip w_1^{-1}\fip^{-1}\fim w_2\fip=
  2\fip(1-w_1^{-1}\vartheta^{-1}w_2\vartheta)\fim .$$
  Since both $\S\P$ and $\S\Q$ are invariant under $2\fim$ and $2\fip$,
  we are left to study the map $1-w_1^{-1}\vartheta^{-1}w_2\vartheta :\S\P\lra
  \S\Q.$ The restriction imposed to $l$ imply that the image of
  $1-w_1^{-1}\vartheta^{-1}w_2\vartheta$  is a split direct summand.\acapo
  It is also clear at this point that the rank of $\Delta$ is precisely
  $l(w_1)+l(w_2)+rk(1-w_1^{-1}\vartheta^{-1}w_2\vartheta).$ But
  $rk(1-w_1^{-1}\vartheta^{-1}w_2\vartheta)=rk(\vartheta w_1-w_2\vartheta)=
  rk(\fip w_1\fim-\fim w_2\fip)$ and we are done.\cvd

\vskip 5mm
{\bf 5.7.} \cor {\sl Let $l$ be a $\vp$-good integer and let $p$ be a point
of  the symplectic
leaf $\Theta$ of {\rm G}.
 Then the dimension of any representation of $\gr$ lying
over $p$ is divisible by $l^{{1\over 2}dim\Theta}$.}\acapo
\vskip 5mm
{\bf 5.8.} As a consequence of (2.2) we have the following isomorphisms
of Hopf algebras
$$\fiobp\simeq \giobm_{op}\ ,\ \ \R_q^0[\B_+]\simeq \Gamma^0(\ben_-)_{op}.$$
Now the algebra $\giobm$ is equal to the algebra
 $\Gamma^0(\ben_-)$ and so we have
the isomorphism of algebras
$$\R_q^\vp[\B_+]\simeq
\R_q^0[\B_+]$$
and similarly for the case $\B_-$.  Then the representations of $\gr$
 over the sets $\X_{(w,1)}$ and $\X_{(1,w)}$ are like in the case $\vp=0$
 (they are  studied in [D-P2]).
In particular there is an isomorphism between the one dimensional
representations of $\gr$ and the points of the Cartan torus T (given
esplicitely
[D-L]).\acapo
\vskip 5mm
\refe{Appendix}
In {\bf 4.} we determined the dimension $d_\vp(w_1,w_2)$ of a symplectic leaf
$\Theta$ contained in $\X_{(w_1,w_2)}$;
$$d_\vp(w_1,w_2)=l(w_1)+l(w_2)+rk(\fip w_1\fim-\fim w_2\fip).$$
This means, of course, that $d_\vp(w_1,w_2)$ is an even integer. Here
we give a direct proof of this fact in the more general context of finite
Coxeter groups. Using the definitions from [H], let $W=<s_1\vir  s_n>$ be
a finite Coxeter group of rank $n$, $\sigma:W\leftarrow GL(V)$, the geometric
representation of $W$, $B$ the $W$-invariant scalar product on $V$, $\Phi$ the
root system of $W$, $l(\cdot)$ the usual lengh function on $W$. We recall
a fact proved in [C] for Weyl groups and which holds with the same proof for
finite Coxeter groups. Each element $w$ of $W$ can be expressed in the
form $w=s_{r_1}\cdots s_{r_k}$, $r_i\in\Phi$, where $s_v$ is the reflection
relative to $v$, if $v$ is any non zero element of $V$.
 Denote by ${\overline l}(w)$
the smallest value of $k$  in any such expression for $w$.
 We get \acapo\vskip 3mm
\lem ${\overline l}(w)=rk(1-w).$\acapo\vskip 3mm
\pf It is Lemma 2 in [C].\cvd\acapo\vskip 3mm
We can now prove\acapo\vskip3mm
{\bf Proposition 1}  {\sl Let $w_1,w_2$ be in $W$. Then
$l(w_1)+l(w_2)+rk(w_1-w_2)$ is even.}\acapo\vskip 3mm
\pf We have
$$rk(w_1-w_2)=rk(1-w_2w_1^{-1})={\overline l}(w_2w_1^{-1})\equiv
l(w_2w_1^{-1})\ mod\ 2.$$
But $l(w_2w_1^{-1})\equiv l(w_2)+l(w_1^{-1})\ mod\ 2$ and
finally $l(w_1^{-1})=l(w_1)$. Hence $l(w_1)+l(w_2)+rk(w_1-w_2)$
$\equiv
l(w_1)+l(w_2)+l(w_2)+l(w_1)\equiv 0\ mod\ 2$.\cvd\vskip 3mm
Suppose now $\vp$ is an endomorphism of $V$ which is skew relative to $B$,
and let $\vartheta$ be the isometry $\fip^{-1}\fim.$ To prove the general
result, we recall that, if $\eta$ is an isometry of $V$ and $r$ is the
rank of $1-\eta$, then $\eta$ can be written as a product of $r+2$ reflections
(cf. [S], where a more precise statement is given). In particular if $\eta=
s_{v_1}\cdots s_{v_k}$, then $rk(1-\eta)\equiv k\ mod\ 2.$\acapo\vskip 3mm
{\bf Proposition 2}  {\sl Let $w_1,w_2$ be in $W$. Then
$l(w_1)+l(w_2)+rk(\fip w_1\fim-\fim w_2\fip)$ is even.}\acapo\vskip 3mm
\pf We have $rk(\fip w_1\fim-\fim w_2\fip )=rk (1-\vartheta w_2\vartheta^{-1}
w_1^{-1}).$ If we write $\vartheta,w_1,w_2$ as products
of $a,a_1,a_2$ reflections
respectively,
 we get from the previous observation that $rk(1-\vartheta w_2\vartheta
^{-1} w_1^{-1})\equiv a+a_2+a+a_1\ mod\ 2.$ Hence
$$rk(1-\vartheta w_2\vartheta^{-1} w_1^{-1})
\equiv rk(1-w_2w_1^{-1})\equiv rk(w_1-w_2)
\ mod\ 2$$
and the result comes from prop.1.\cvd
\refe{Acknowledgements}
The authors are grateful to  C. De Concini for  helpful discussions
and enlightening
explanations.

\refe{References}
\item\item{{[C]}} Carter, R.W.: Conjugacy Classes in the Weyl Group.
Compositio Math. {\bf 25},1-59 (1972)\acapo
\item\item{{[C-V]}}  Costantini, M., Varagnolo, M.: Quantum Double and
Multiparameter Quantum Groups. to appear in  Comm.Algebra\acapo
\item\item{{[D-K-P1]}} De Concini, C., Kac, V.G., Procesi, C.:
Quantum Coadjoint Action.  J.of AMS {\bf 5},151-189 (1992)\acapo
\item\item{{[D-K-P2]}} De Concini, C., Kac, V.G., Procesi, C.:
Some Quantum Analogues of Solvable Groups. preprint\acapo
\item\item{{[D-L]}}  De Concini, C., Lyubashenko, V.: Quantum Function
Algebra at roots of 1. to appear in  Adv.Math.\acapo
\item\item{{[D-P1]}}  De Concini, C., Procesi, C.: Quantum Groups.
 preprint\acapo
\item\item{{[D-P2]}}  De Concini, C., Procesi, C.: Quantum Schubert Cells
and Representations at Roots of 1. preprint\acapo
\item\item{{[D1]}}  Drinfel'd, V.G.: Hopf Algebras and the Quantum
Yang-Baxter Equation.  Soviet.Math.Dokl. {\bf 32},254-258 (1985)\acapo
\item\item{{[D2]}} Drinfel'd, V.G.: Quantum Groups, Proceedings of the
ICM, AMS, Providence. R.I. {\bf 1},798-820 (1987)\acapo
\item\item{{[H]}} Humpreys, J.E.: Reflection Groups and Coxeter Groups.
 London : Cambridge University Press 1990\acapo
\item\item{{[H-L1]}} Hodge, T.J., Levasseur, T.:
Primitive ideals of $\CI_q[SL(3)]$. Comm.Math.Phys.
 {\bf 156},581-605 (1993)\acapo
\item\item{{[H-L2]}} Hodge, T.J., Levasseur, T.:
Primitive ideals of $\CI_q[SL(n)]$. J. Algebra, to appear\acapo
\item\item{{[H-L3]}} Hodge, T.J., Levasseur, T.:
Primitive ideals of $\CI_q[G]$. preprint\acapo
\item\item{{[J]}} Jimbo, M.: A q-difference analog of $U(\gen)$ and
the Yang-Baxter Equation.  Lett.Math.Phys. {\bf 10},63-69 (1985)\acapo
\item\item{{[Jo]}} Joseph, A.: On the Prime and Primitive Spectra
of the Algebra of Functions on a Quantum group. preprint\acapo
\item\item{{[L-S1]}}  Levendorskii, Ya.S.Soibelman, Ya.S.: Quantum
Weyl Group and Multiplicative Formula for the R-Matrix of a Simple
Lie Algebra. Funk. Anal. Pri. {\bf 25},73-76 (1991)\acapo
\item\item{{[L-S2]}}  Levendorskii, Ya.S.Soibelman, Ya.S.: Algebras of
Functions on Compact Quantum Groups, Schubert Cells and Quantum Tori.
 Commun.Math.Phys. {\bf 139},171-181 (1991)\acapo
 \item\item{{[L-W]}} Lu, J.H., Weinstein, A.: Poisson Lie Groups,
 Dressing Transformations and Bruhat Decompositions. J.Diff.
 Geometry {\bf 31},501-526 (1990)\acapo
 \item\item{{[L1]}} Lusztig, G.: Quantum Deformations of Certain
 Simple Modules over Enveloping Algebras.
 Adv. in Math {\bf70},237-249 (1988)\acapo
\item\item{{[L2]}} Lusztig, G.: Quantum Groups at Roots of 1.
Geom.Dedicata {\bf 35},89-113 (1990)\acapo
\item\item{{[R]}} Reshetikhin, N.: Multiparameter Quantum Groups and
Twisted Quasitriangular Hopf Algebras.  Lett.Math.Phys. {\bf 20},331-335
(1990)\acapo
\item\item{{[S]}} Scherk, P.: On the Decomposition of Ortogonalities
into Symmetries. In : Proceeding of A.M.S., vol.1,481-491 (1950)\acapo
\item\item{{[T]}} Takeuchi, M.: Matched Pairs of Groups and Bismash
Products of Hopf Algebras.  Comm.Algebra {\bf 9},841-882 (1981)\acapo

\bye